\documentclass[reprint,preprintnumbers, amsmath,amssymb,aps,pra]{revtex4-1}
\usepackage{cancel}
\usepackage{lipsum}
\usepackage{graphicx}
\usepackage{dcolumn}
\usepackage{bm}
\usepackage[dvipsnames,usenames]{xcolor}
\usepackage{amsmath}
\usepackage{amssymb} 
\usepackage{hyperref}
\usepackage[qm,braket]{qcircuit}
\usepackage{bbold}
\usepackage{MnSymbol}
\usepackage{algorithm}
\usepackage[noend]{algpseudocode}

\usepackage{mathtools}

\begin{document}
\title{Spectral density estimation with the Gaussian Integral Transform}
\author{A. Roggero}
\email{roggero@uw.edu}
\affiliation{Institute for Nuclear Theory, University of Washington, 
Seattle, WA 98195, USA}

\preprint{INT-PUB-20-014 }

\date{\today}
\begin{abstract}
The spectral density operator $\hat{\rho}(\omega)=\delta(\omega-\hat{H})$ plays a central role in linear response theory as its expectation value, the dynamical response function, can be used to compute scattering cross-sections. In this work, we describe a near optimal quantum algorithm providing an approximation to the spectral density with energy resolution $\Delta$ and error $\epsilon$ using $\mathcal{O}\left(\sqrt{\log\left(1/\epsilon\right)\left(\log\left(1/\Delta\right)+\log\left(1/\epsilon\right)\right)}/\Delta\right)$ operations. This is achieved without using expensive approximations to the time-evolution operator but exploiting instead qubitization to implement an approximate Gaussian Integral Transform (GIT) of the spectral density. We also describe appropriate error metrics to assess the quality of spectral function approximations more generally.
\end{abstract}
\maketitle

Since the first seminal works of Feynman~\cite{Feynman1982} and Lloyd~\cite{Lloyd96}, quantum computing has been recognized as a possible avenue to explore quantum dynamics of strongly correlated many-body systems beyond what is possible with classical computational tools. Recent progress in hamiltonian simulation algorithms~\cite{Berry2015,Low2017,Low2019,Babbush2019} has allowed a dramatic reduction of the computational cost for applications as diverse as computing out-of-equilibrium dynamics~\cite{Lamm2018}, exclusive scattering cross-sections~\cite{RoggeroLR,RoggeroNUQC} and ground state energy estimation~\cite{Childs2018}.
Most of the proposed algorithms still require a number of gates too large for possible applications on NISQ devices~\cite{Preskill2018} and more work is required to bring these costs down (see eg.~\cite{RoggeroNUQC} for a recent analysis of the requirements for neutrino-nucleus scattering). 

In the same spirit of the recent work by Somma~\cite{Somma2019}, we propose in this work a new quantum algorithm with near optimal computational cost (in terms of oracle calls) to study the problem of spectral density estimation. In particular, given an hermitian operator $\hat{O}$, the goal of this work is to obtain an efficient algorithm to approximate the spectral density operator $\hat{\rho}(\omega)=\delta\left(\omega-\hat{O}\right)$, with $\delta$ the Dirac delta function. Using the eigenstates $\ket{k}$ of the operator $\hat{O}$ we have the following spectral representation
\begin{equation}
\hat{\rho}(\omega)=\sum_k^\Gamma \delta\left(\omega-O_k\right)\rvert k\rangle\langle k\lvert\;.
\end{equation}
with $O_k$ the eigenvalue for $\ket{k}$, and $\Gamma$ the total number of eigenvalues. Without loss of generality, we will consider normalized operators $\hat{O}$ with $\|\hat{O}\|\leq1$ so that the spectrum is contained in the interval $[-1,1]$.

One of the most popular applications of the spectral density operator is in the theory of linear response where it is directly connected with the {\it dynamical response function} $S(\omega)$. More precisely, given a state vector $\ket{\Psi}$ we can define the following response function
\begin{equation}
\label{eq:resp}
\begin{split}
S(\omega) &= \langle\Psi\lvert\hat{\rho}(\omega)\rvert\Psi\rangle=\sum_k^\Gamma |\langle\Psi|k\rangle|^2 \delta\left(\omega-O_k\right)\;.
\end{split}
\end{equation}
The response function can be used to compute, among other things, the energy resolved inclusive cross section for a scattering process that maps an initial state $\ket{\Phi_0}$ to the final state $\ket{\Psi}=\hat{Q}\rvert\Phi_0\rangle$ trough the action of the (possibly non unitary) vertex operator $\hat{Q}$. In this case, the relevant operator $\hat{O}$ coincides with the Hamiltonian of the physical system, and for this reason we will often call it's eigenvalues "frequencies". The technique we describe here is however applicable to any hermitian operator.

The approach we follow in this work is to consider approximations to response function obtained trough an integral transform of the type
\begin{equation}
\label{eq:inttrasf}
\begin{split}
\Phi_K(\nu) &= \int d\omega K(\nu,\omega)S(\omega) = \sum_k^\Gamma |\langle\Psi|k\rangle|^2 K(\nu,O_k)\;.
\end{split}
\end{equation}
The integral kernel $K(\nu,\omega)$ that defines the transform can also be used directly as an approximation to the spectral density operator: $\hat{\rho}_K(\nu)=K\left(\nu,\hat{O}\right)\approx\hat{\rho}(\omega=\nu)$.
For this to be a good approximation, the kernel function should be chosen as a finite width representation of the Dirac delta-function. 

We note that the approach of computing response functions by a direct inversion of integral transforms like Eq.~\eqref{eq:inttrasf} is a common strategy in many-body physics. In Quantum Monte Carlo calculations, for instance, it is common to consider the Laplace kernel $K(\omega,\nu)=\exp(-\nu\omega)$ due to it's connection with euclidean path integrals (see eg.~\cite{Carlson92,Ceperley}), but other alternatives such as the Sumudu~\cite{Roggero2013} and Lorentz~\cite{Efros1994,Efros2007,Bacca2013} transforms have also been considered in the past. The main difficulty encountered by these methods is the problem that, for any compact kernel function, the inversion of the integral transform is a numerically ill-posed problem: any errors in the estimate of $\Phi_K$ will get exponentially amplified by the inversion procedure (see eg.~\cite{Glockle2009,Barnea2010}). In this work we avoid the problem by using directly the integral transform $\Phi_K(\omega)$ as the approximate reconstruction of the original signal $S(\omega)$. We note at this point that the idea of using directly the integral transform to extract physical informations has been explored already in the past with great success. For example, in~\cite{miorelli2016} the dipole polarizability $\alpha_D$ of $^{22}O$ was computed using the Coupled Cluster method and using a direct mapping between $\alpha_D$ and a Lorentz Integral Transform of the response function and in~\cite{roggero2016} the contribution of impurity scattering in the thermal conductivity in the outer crust of neutron stars was successfully extracted by mapping it into features of the Laplace Transform of the response. 

A possible future extension of our work would be to consider approximate inversion schemes like the Maximum Entropy Method~\cite{Gubernatis91} to try and reduce the computational cost of the quantum algorithm at the possible expense of introducing an uncontrollable error.

The paper is organized as follows, in Sec.~\ref{sec:intro} we first provide a detailed description of the error metrics we use to judge the quality of the approximation in Eq.~\eqref{eq:inttrasf} and in Sec.~\ref{sec:cmp} we summarize the main results of the paper and compare them to the recent work from Ref.~\cite{Somma2019} which can be understood as a particular instance of the method we propose. We also provide an argument for the near optimatility of both techniques. We then present two integral kernels: the Fejer kernel naturally generated using Quantum Phase Estimation~\cite{RoggeroLR,RoggeroNUQC} in Sec.~\ref{sec:fejer} and the Gaussian kernel which allows to achieve near optimal scaling of the computational cost in Sec.~\ref{sec:git}. We also provide a pseudocode implementation in Appendix~\ref{sec:pcode}. We conclude in Sec.~\ref{sec:conclusions} providing a summary of our findings and proposing possible avenues for future improvements.

\section{Definitions and comparison to previous work}
\label{sec:intro}

In order to precisely quantify the accuracy of the approximation procedure presented in the introduction above, and connect with recent work on quantum algorithms exploring similar problems~\cite{Novo2019,Somma2019}, we now introduce the following definitions:
\begin{itemize}
    \item we will call an integral kernel {\it $\Sigma$-accurate} with {\it resolution $\Delta$} if the following condition holds
    \begin{equation}
    \label{eq:sigma_approx}
    \sup_{\omega_0\in[-1,1]}\sumint_{\omega_0-\Delta}^{\omega_0+\Delta} d\nu K(\nu,\omega_0)\geq1-\Sigma\;,
    \end{equation}
    where the symbol $\sumint$ indicates: an integral when the transformed variable $\nu$ is defined over a continuous interval or a sum if $\nu$ is defined on a discrete set.
    
    \item we will call a distribution $\widetilde{\Phi}(\omega)$ a {\it $\beta$-approximation} to the true distribution $\Phi(\omega)$ with confidence $1-\eta_\beta$ if the total variation is bounded as
    \begin{equation}
    \label{eq:totvar}
    \delta_V(\Phi,\widetilde{\Phi})\coloneqq\sup_{\omega\in[-1,1]}\left|\Phi(\omega)-\widetilde{\Phi}(\omega)\right|\leq\beta\;.
    \end{equation}
    with probability $P>1-\eta_\beta$.
    \item if the estimator $\widetilde{\Phi}_K$ is obtained as a {\it $\beta$-approximation} with {\it confidence $1-\eta_\beta$} of a {\it $\Sigma$-accurate} integral transform $\Phi_K$ of the response function $S(\omega)$ with {\it resolution $\Delta$} we will call it a $(\Sigma,\Delta,\beta,\eta_\beta)$-approximation to the response $S$.
\end{itemize}

These definitions are similar to those introduced in the recent work Ref.~\cite{Novo2019}. In particular, the first definition is similar in spirit to, but more stringent than, the condition of having {\it resolution $\Delta$} and {\it confidence $\eta=(1-\Sigma)$} (Definition 1 of~\cite{Novo2019}) while the second condition is equivalent to the {\it $\beta$-approximation} (Definition 3 of~\cite{Novo2019}). 

The reason for these definitions, and the mild departure from those introduced in Ref.~\cite{Novo2019}, is rooted in the fact that for physics application we are ultimately interested in frequency observables of the form
\begin{equation}
\label{eq:obsofs}
Q(S,f) = \int_{-1}^1d\omega S(\omega)f(\omega)
\end{equation}
for some bounded function $f$. If we estimate the observable $Q$ using a $(\Sigma,\Delta,\beta,\eta_\beta)$-approximation $\widetilde{\Phi}_K$ we have in fact, with confidence $1-\eta_\beta$, the following bound 
\begin{equation}
\label{eq:maxerr}
\bigg|Q(S,f)-Q(\widetilde{\Phi}_K,f)\bigg|\leq f^\Delta_{max} + 2f_{max}\Sigma+\beta f_{int}\;,
\end{equation}
where we have defined the quantities
\begin{equation}
f_{max} = \sup_{\omega\in[-1,1]}\left|f(\omega)\right|\quad f_{int} = \int_{-1}^1d\omega \left| f(\omega)\right|\leq2f_{max}\;,
\end{equation}
and the upperbound on the maximum variation
\begin{equation}
f^{\Delta}_{max} = \sup_{\omega\in[-1,1]} \sup_{x\in[-\Delta,\Delta]} \left|f(\omega+x)-f(\omega)\right|\;.
\end{equation}
A full derivation of this is provided in Appendix~\ref{app:proof_obs_bound}.

At this point it is important to point out another difference with Ref.~\cite{Novo2019}. In our work, the second error metric $\beta$ captures both the statistical error coming from estimating the distribution $\widetilde{\Phi_K}(\omega)$ with a finite number of samples, but also the possible systematic error coming from using an approximation of the quantum circuit needed to obtain the desired integral transform $\Phi_K(\omega)$. In this sense being {\it $\Sigma$-accurate} with {\it resolution $\Delta$} is a property of the kernel function $K(\nu,\omega)$, while being a {\it $\beta$-approximation} with {\it confidence $1-\eta_\beta$} is a property that characterizes the implementation of the algorithm that generates the desired integral transform.

\subsection{Comparison to previous work}
\label{sec:cmp}

\begin{table*}[t]
\begin{tabular}{c|c|c}
Method & Number of calls to $W_Q$ & Total number of samples \\ \hline
&&\\
TSA & $\mathcal{O}\left(\frac{1}{\Delta}\log\left(\frac{1}{\varepsilon}\right)^2\right)$ &$\mathcal{O}\left(\frac{1}{\Delta^3\varepsilon^2}\log\left(\frac{1}{\varepsilon}\right)^6\log\left(\frac{1}{\eta_\beta}\right)\right)$ \\
&&\\
Fejer & $\mathcal{O}\left(\frac{1}{\Delta\varepsilon}\right)$ & $\mathcal{O}\left(\frac{1}{\varepsilon^2}\log\left(\frac{1}{\eta_\beta}\right)\right)$ \\
&&\\
GIT & $\mathcal{O}\left(\frac{1}{\Delta}\sqrt{\log\left(\frac{1}{\varepsilon}\right)\log\left(\frac{1}{\Delta\varepsilon}\right)}\right)$ & $\mathcal{O}\left(\frac{1}{\Delta^3\varepsilon^2}\left(\log\left(\frac{1}{\varepsilon}\right)\log\left(\frac{1}{\Delta\varepsilon}\right)\right)^{3/2}\log\left(\frac{1}{\eta_\beta}\right)\right)$\\
\end{tabular}
\caption{Comparison of the computational cost required to obtain a $(\varepsilon,\Delta,\varepsilon,\eta_\beta)$-approximation to the response function using: the time series analysis (TSA) method of Ref.~\cite{Somma2019}, the Fejer based methods from Refs.~\cite{RoggeroLR,RoggeroNUQC} and the new GIT-based method proposed in this work. See also Appendix.~\ref{sec:pcode} for the asymptotic scaling in a different limit.\label{tab:summary}}
\end{table*}

The approximation problem we are trying to solve is very similar to the Quantum Eigenvalue Estimation Problem (QEEP) considered in Ref.~\cite{Somma2019}. In this section we will anticipate the main results of our work and provide a comparison with the Time-Series Analysis algorithm (TSA) proposed in Ref.~\cite{Somma2019}. In particular, we will compare the computational cost in terms of the number $M$ of oracle calls to a base unitary $W_Q$ and the total number $N_S$ of samples needed to generate a $(\Sigma,\Delta,\beta,\eta_\beta)$-approximation to the spectral function $S(\omega)$. In order to simplify the comparison, we will consider here the limit $\Sigma=\beta=\varepsilon$ which is sensible given the definition in Eq.~\eqref{eq:maxerr}. Detailed results for the more general case will be provided in the sections below.

The TSA approach from Ref.~\cite{Somma2019} starts by decomposing the frequency domain into $N$ disjoint intervals of size $2\Delta$ and then obtaining the response in each of these bins using the Fourier expansion of the bump-function. In light of the definitions provided above, this can be understood as using an integral transform with kernel function given by the approximate frequency comb
\begin{equation}
K(\nu_j,\omega) = \sum_{j=1}^{N} f_b(\nu_j,\omega)\;,
\end{equation}
where $\nu_j$ is the central value of the $j$-th frequency bin and the function $f_b$ is obtained from bump-functions and has support on $[\nu_j-\Delta,\nu_j+\Delta]$ only. Due to this property, it is straightforward to see that this kernel allows to achieve accuracies $\Sigma=0$ in Eq.~\eqref{eq:sigma_approx}. Note however that using $\Sigma\ll\beta$ will not help reduce the final error in Eq.~\eqref{eq:maxerr} (unless $f_{max}\gg f_{int}$), and in fact here we only require them to be both equal to $\varepsilon$. 

The TSA algorithm requires to apply the (controlled) time evolution operator $U_O(t)=\exp(-it\hat{O})$ for a maximum time $t_{max}$ scaling as (see Appendix A of~\cite{Somma2019})
\begin{equation}
\label{eq:maxtaurolando}
t_{max}=\mathcal{O}\left(\frac{1}{\Delta}\log\left(\frac{1}{\varepsilon}\right)^2\right)
\end{equation}
together with a total number of samples scaling as
\begin{equation}
\label{eq:nsrolando}
N_S=\mathcal{O}\left(\frac{1}{\Delta^3\varepsilon^2}\log\left(\frac{1}{\varepsilon}\right)^6\log\left(\frac{1}{\eta_\beta}\right)\right)
\end{equation}
in order to achieve a $(\Sigma=0,\Delta,\varepsilon,\eta_\beta)$-approximation. Note that if we require the final approximation over $N$ frequency to have total error less than $\varepsilon$ (as done in~\cite{Somma2019}) the $\varepsilon$-dependent logarithmic terms above will include an additional $1/\Delta$ like GIT (see also Appendix~\ref{sec:pcode}).

In order to compare these asymptotic scaling with the bounds provided in our work, while at the same time account for the unavailable bound on the time evolution error for the TSA method, we consider here the situation where we use the optimal time evolution scheme of Ref.~\cite{Low2017} (which is based on qubitization~\cite{Low2019}) and neglect the mild overhead needed to improve the precision to the desired level. Using this implementation, the number of applications of the qubiterate unitary $W_Q$ (see Sec.~\ref{sec:fejer} for more details) is simply $M=\mathcal{O}\left(t_{max}\right)$.

In this work we consider two different integral transforms. The first is associated with the Fejer kernel that is naturally produced by using the Quantum Phase Estimation (QPE) algorithm~\cite{Cleve1998} to approximate the response as described in Refs.~\cite{RoggeroLR,RoggeroNUQC}. The second is a Gaussian Integral Transform (GIT) obtained using the connection between quantum walks and Chebyschev polynomials~\cite{Childs2017}. We will analyze these integral transform in detail in the next sections and anticipate here the main results.

Due to the choice $\beta=\Sigma=\varepsilon$, both the standard Fejer method of Ref.~\cite{RoggeroLR} and the qubitization-based variant from Ref.~\cite{RoggeroNUQC} have the same asymptotic scaling. We will anticipate here results for the latter, which can produce a $(\varepsilon,\Delta,\varepsilon,\eta_\beta)$-approximation using 
\begin{equation}
\label{eq:callsfqub}
M = \mathcal{O}\left(\frac{1}{\Delta\varepsilon}\right)\quad\text{and}\quad N_S=\mathcal{O}\left( \frac{1}{\varepsilon^2}\log\left(\frac{1}{\eta_\beta}\right)\right)\;.
\end{equation}
Even tough the sample complexity is greatly reduced, for small target errors $\varepsilon$ the gate count of this scheme will be larger than the estimate obtained from Eq.~\eqref{eq:maxtaurolando}. Despite this, as described in detail in Sec.~\ref{sec:fejer}, this scheme could still be beneficial as it avoids performing an approximation to the time-evolution operator.

As we will show in more detail in Sec.~\ref{sec:git}, using the GIT provides a considerable reduction of the quantum computational cost (ie. the gate count) compared to both methods described above. This comes at the cost of requiring a larger number of measurements $N_S$ than the Fejer-based methods, but still less or comparable to Eq.~\eqref{eq:nsrolando}. In particular, we will find that a $(\varepsilon,\Delta,\varepsilon,\eta_\beta)$-approximation to the response function requires only
\begin{equation}
\label{eq:qcmplxGIT}
M=\mathcal{O}\left(\frac{1}{\Delta}\sqrt{\log\left(\frac{1}{\varepsilon}\right)\log\left(\frac{1}{\Delta\varepsilon}\right)}\right)
\end{equation}
calls to the qubiterate unitary $W_Q$, together with 
\begin{equation}
N_S = \mathcal{O}\left(\frac{1}{\Delta^3\varepsilon^2}\left(\log\left(\frac{1}{\varepsilon}\right)\log\left(\frac{1}{\Delta\varepsilon}\right)\right)^{3/2}\log\left(\frac{1}{\eta_\beta}\right)\right)
\end{equation}
samples. We summarize these estimates in Tab.~\ref{tab:summary} and provide a pseudocode implementation in Appendix~\ref{sec:pcode}. 

That the quantum query complexity Eq.~\eqref{eq:qcmplxGIT} is almost optimal can be seen by looking at our approach as a technique to estimate the ground state energy of some hamiltonian as in Ref.~\cite{Lin20}. In particular, optimality can be shown by considering: an hamiltonian with spectral gap $(\omega_1-\omega_0)>2\Delta$, an initial state $\ket{\Psi}$ with an overlap on the ground state state $|\langle\Psi\vert0\rangle|\geq\varepsilon$ and ask for an approximation of the ground state energy with probability $P>1-\varepsilon$ and confidence $1-\eta_\beta$. Using the results from Ref.~\cite{Lin20} (Lemmas 3 and 5 and Theorems 8 and 9) we know that this requires at least $M=\mathcal{O}\left(1/\Delta\log(1/\varepsilon)\right)$ oracle calls to $W_Q$. We can also solve this problem by considering a $(\varepsilon,\Delta,\varepsilon,\eta_\beta)$-approximation $\widetilde{\Phi}_K$ to the response $S_2(\omega)=\langle\Psi\lvert\hat{\rho}(\omega)\rvert\Psi\rangle$. Our result is then only a factor $\mathcal{O}(\sqrt{\log(1/\varepsilon)\log(1/\Delta)})$ away from the optimal result and provides a quadratic speedup in the logarithmic factors compared to the TSA scheme of Ref.~\cite{Somma2019}. 

\section{Fejer kernel}
\label{sec:fejer}
The standard Quantum Phase Estimation (QPE) algorithm~\cite{Cleve1998,Berry2000} uses $n$ applications of the (controlled) time evolution unitary $U_O(t)=\exp\left(-it\hat{O}\right)$ and $N=2^n$ ancilla qubits to approximately diagonalize the "hamiltonian" operator $\hat{O}$~\footnote{If needed, the number of ancilla qubits can be reduced to just $1$ using iterative schemes.}. As we proposed in Ref.~\cite{RoggeroLR}, this technique can be used to perform an integral transform generated by a rescaled Fejer kernel
\begin{equation}
\label{eq:fejer}
K_F(\sigma_q,\omega,N) = \frac{1}{N^2} \frac{\sin^2\left(N\pi(\sigma_q-\omega)/2\right)}{\sin^2\left(\pi(\sigma_q-\omega)/2\right)}
\end{equation}
where the discrete frequencies $\sigma_q$ are defined on a grid with $N$ points: $\sigma_k = \left(2k/N\right)-1$ for $k=\{0,\dots,N-1\}$. The integer parameter $N>1$ controls the maximum propagation time $t_{max}$ used in QPE as $t_{max}=\pi N$. 
In this case, ensuring the resulting integral transform $\Phi_F(\omega)$ is {\it $\Sigma$-accurate} with {\it resolution $\Delta$} is equivalent to requiring the probability of measuring a phase $\sigma_k$ with error more than $\Delta$ to be less than $\Sigma$. This probability can be bounded using standard techniques (see eg. 5.2.1 of~\cite{NieChu}) as
\begin{equation}
\begin{split}
P\left(\left|k-\frac{N}{2}(\omega+1)\right|>\frac{N\Delta}{2}\right)\leq\frac{1}{N\Delta-2}\;,
\end{split}
\end{equation}
which then implies we can take the closest power of $2$ of
\begin{equation}
N\geq \frac{1}{\Delta}\left(\frac{1}{\Sigma}+2\right)\;,
\end{equation}
in order to satisfy Eq.~\eqref{eq:sigma_approx}. The dependence on the resolution $\Delta$ is already optimal and the constant factors could be improved using optimized preparations of the ancilla register~\cite{Berry2000}. The scaling with the error $\Sigma$ instead could be improved to $N=\mathcal{O}(\log(1/\Sigma)1/\Delta)$ in the special situation where the signal $S(\omega)$ is composed by a single frequency mode by using schemes like Kitaev's original algorithm~\cite{KitaevQPE} or the more efficient IPEA~\cite{Wiebe2016}. In the general case where the number of modes in the response of Eq.~\eqref{eq:resp} satisfies $\Gamma\gg1$, this is is not in general possible (see eg.~\cite{OBrien2019}). We can now use $N_S=\mathcal{O}\left(1/\beta^2\right)$ samples to produce the {\it $\beta$-approximate} estimator $\widetilde{\Phi_F}$ by collecting an histogram of the measured frequencies. More precisely, using Hoeffding's inequality~\cite{Hoeffding} we find it sufficient to take
\begin{equation}
\label{eq:nsamplesfA}
N_S=\frac{1}{2\beta^2}\log\left(\frac{2}{\eta_\beta}\right)\;,
\end{equation}
with $\eta_\beta$ the confidence of the $\beta$-approximation. 

In general, the time-evolution operator $U_O(t)$ needs to be approximated with additive error $\delta_t$, using available quantum operations, and a proper consideration of this approximation error is critical for a fair assessment of the overall computational cost. As discussed in Sec.~\ref{sec:intro}, we will consider these errors as contributions to the total variation Eq.~\eqref{eq:totvar} which define the {\it $\beta$-approximation}. In particular, if we denote by $\Phi^{e}_F(\omega)$ the transform obtained by using the approximate time-evolution unitary and $\widetilde{\Phi^{e}_F}(\omega)$ it's finite population estimator, we have
\begin{equation}
\begin{split}
\delta_V(\Phi_F,\widetilde{\Phi^{e}_F})
&\leq \delta_V(\Phi_F,\Phi^{e}_F) + \delta_V(\Phi^{e}_F,\widetilde{\Phi^{e}_F})\;.
\end{split}
\end{equation}
The second term measures statistical fluctuations and can be dealt with using again the Hoeffding bound, for the first term instead in Appendix~\ref{app:proof_faulty_fejer} we show that
\begin{equation}
\delta_V(\Phi_F,\Phi^{e}_F)\leq \log_2(N)\delta_t\;,
\end{equation}
with $\delta_t$ an upperbound to the approximation error of the time-evolution operator for times up to $t_{max}=\pi N$. The finite population estimator of the approximate Fejer transform is then {\it $\beta$-accurate} with confidence $\eta_\beta$ if
\begin{equation}
N_S=\frac{2}{\beta^2}\log\left(\frac{2}{\eta_\beta}\right)\quad\delta_t\leq\frac{\beta}{2\log_2(N)}\;.
\end{equation}
Using optimal scaling algorithms for time evolution like Quantum Signal Processing~\cite{Low2017}, the total gate count is
\begin{equation}
\label{eq:MscalingfejerA}
\begin{split}
M
&=\mathcal{O}\left(\frac{1}{\Delta\Sigma}+\log\left(\frac{1}{\beta}\right)\right)\;,
\end{split}
\end{equation}
in terms of oracle queries to the a basic quantum subroutine: the qubiterate $W_Q$. This unitary is defined as  
\begin{equation}
\label{eq:qubiterate}
W_Q=\exp\left(i\widetilde{Y}\arccos\left(\hat{O}\right)\right)
\end{equation}
where $\widetilde{Y}$ is an isometry defined over a two dimensional space for each energy eigenvalue (see~\cite{Low2017,Low2019} and the Appendix of Ref.~\cite{Childs2018} for a complete derivation). The most important property of $W_Q$ for our purposes is that it can be implemented exactly and efficiently. It is important to point out that short-time approximation methods based on the Trotter-Suzuki~\cite{Suzuki91} expansion are not able to achieve the optimal scaling in Eq.~\eqref{eq:MscalingfejerA}.

A slight modification to this scheme, with the same scaling but possibly greatly reduced prefactors, can be obtained by applying the QPE algorithm directly on the qubiterate $W_Q$ (see Ref.~\cite{RoggeroNUQC}). One can easily show that this leads to a modified Fejer kernel given by
\begin{equation}
K_{FQ}(\sigma_q,\omega,N) = \frac{K_F(\sigma_q,\theta_\omega,N)+K_F(\sigma_q,-\theta_\omega,N)}{2}
\end{equation}
where we have defined $\cos(\theta_\omega)=\omega$. In order to distinguish the two peaks at $\pm\theta_\omega$ we can shift and rescale the excitation operator $\hat{O}$ so that its spectrum lies in $[0,1]$ only. The needed resolution in this transformed space (apart from the trivial factor of $1/2$ coming from the rescaling) will need to satisfy
\begin{equation}
\left|\cos\left(\theta_\omega\pm\Delta_\theta\right)-\cos\left(\theta_\omega\right)\right|\leq\frac{\Delta}{2}\;
\end{equation}
which amounts to require $\Delta_\theta\leq\sqrt{1+\Delta}-1$.
We then find that, in order to obtain a $(\Sigma,\Delta,\beta,\eta_\beta)$-approximation to the response function, the qubitization based Fejer transform of Ref.~\cite{RoggeroNUQC} requires the closest power of $2$
\begin{equation}
M\geq\frac{2}{\Delta_\theta}\left(\frac{1}{\Sigma}+2\right)\gtrsim\frac{4}{\Delta}\left(\frac{1}{\Sigma}+2\right)\;,
\end{equation}
black box invocations of the qubiterate $W_Q$~\footnote{note the additional factor of two coming from the need in QPE to perform $\log_2(M)$ controlled evolutions} together with the same number of samples reported in Eq.~\eqref{eq:nsamplesfA}. 
Despite the possible slight increase in oracle calls with respect to the time-evolution based Fejer scheme presented before, by avoiding the overhead in approximating the time evolution operator $U_O(t)$ we expect this strategy to require shorter circuit depths and at the same time less cumbersome controlled operations.  

In the next section we consider algorithms with exponentially better dependence on $\Sigma$.

\section{Gaussian kernel}
\label{sec:git}

We consider now a Gaussian Integral Transform (GIT) defined trough the following kernel function
\begin{equation}
\label{eq:git_kernel}
K_G(\sigma,\omega,\Lambda)=\frac{1}{\sqrt{2\pi}\Lambda}\exp\left(-\frac{(\sigma-\omega)^2}{2\Lambda^2}\right)
\end{equation}
where $\Lambda>0$ controls the resolution, and the transformed frequency $\sigma$ is defined over the whole real line~\footnote{We can also add an additional normalization factor $\mathcal{N}(\sigma,\Lambda)$ that could be used to keep the kernel normalized (and hence maintain the validity of sum-rules) while restricting the values of $\sigma$ to lie in the range $[-1,1]$ as the frequency. We didn't find any significant advantage in doing this we will take $\sigma$ to be defined over the full real line}. The first step is to determine the conditions for which the approximate response obtained using the GIT is $\Sigma$-accurate with resolution $\Delta$. Using the translational invariance of the kernel $K_G(\sigma,\omega,\Lambda)$ for $\sigma\in\mathbb{R}$, we can rewrite the condition Eq.~\eqref{eq:sigma_approx} in terms of the error function as
\begin{equation}
\frac{1}{\sqrt{2\pi}\Lambda}\int_{-\Delta}^{\Delta}d\sigma\exp\left(-\frac{\sigma^2}{2\Lambda^2}\right)=\text{erf}\left(\frac{\Delta}{\sqrt{2}\Lambda}\right)\geq1-\Sigma\;.
\end{equation}
A sufficient condition for this to hold is to choose the kernel resolution $\Lambda$ according to
\begin{equation}
\Lambda \leq \frac{\Delta}{\sqrt{2\log(1/\Sigma)}}\;.
\end{equation}
We now move on to find the condition for the GIT to be $\beta$-approximate with confidence $\eta_\beta$ according to Eq.~\eqref{eq:totvar}. As we mentioned in the introduction, this property is directly connected with the specific implementation of the GIT, and the way we estimate it. Here we consider an approximate implementation of the Gaussian kernel $K_G(\sigma,\omega,\Lambda)$ using an expansion in a set of orthogonal polynomials. Due to it's direct connection with quantum walks~\cite{Childs2017,Subramanian2019} and the qubitization method~\cite{Low2019}, we consider here the basis spanned by the Chebyshev polynomials $T_k$. In particular one can show that, if we indicate with $\ket{G}$ the flag state in the ancilla register used for the block encoding of the excitation operator $\langle G\lvert W_Q\rvert G\rangle=\hat{O}$  and entering in the definition of the qubiterate $W_Q$, we have (see the proof of Lemma 16 of Ref.~\cite{Childs2017} and Appendix D of Ref.~\cite{Subramanian2019})
\begin{equation}
W_Q^k \rvert G\rangle\otimes\rvert\Psi\rangle= \rvert G\rangle\otimes T_k\left(\hat{O}\right)\rvert\Psi\rangle + \ket{\Phi^\perp}\;,
\end{equation}
with $\ket{\Psi}$ the initial state that defines the response function $S(\omega)$ in Eq.~\eqref{eq:resp} and $\ket{\Phi^\perp}$ not normalized and orthogonal to the flag state $\ket{G}$. The expectation value of the $k$-the Chebyshev polynomial can then be obtained as
\begin{equation}
\label{eq:exptofk}
\langle\Psi\lvert T_k\left(\hat{O}\right)\rvert\Psi\rangle = \langle\Psi_G\lvert W_Q^k\rvert\Psi_G\rangle\;,
\end{equation}
where for convenience we have defined $\ket{\Psi_G}\coloneqq\rvert G\rangle\otimes\rvert\Psi\rangle$.
Note that this procedure is deterministic since we are computing a single polynomial at a time.
An exact representation for the GIT can be obtained considering first the series expansion of the Gaussian function
\begin{equation}
\exp\left(-\frac{\omega^2}{2\Lambda^2}\right)=\sum_{k=0}^\infty a_k(\Lambda) T_k\left(\omega\right)
\end{equation}
and then expanding the integral kernel as
\begin{equation}
\label{eq:gaussexp}
\begin{split}
K_G(\sigma,\omega,\Lambda)&=\frac{1}{\sqrt{2\pi}\Lambda} \sum_{k=0}^\infty a_k\left(\frac{\Lambda}{2}\right) T_k\left(\frac{\sigma-\omega}{2}\right)\\
&=\sum_{k=0}^\infty c_k\left(\Lambda,\sigma\right) T_k\left(\omega\right)\;.
\end{split}
\end{equation}
The step leading to the second line is necessary to be able to implement the GIT using qubitization, and the new expansion coefficients $c_k$ can be obtained from the bare $a_k$ and polynomials in $\sigma$. Explicit expressions for these coefficients can be found in Eq.~\eqref{eq:ccoeff} of Appendix~\ref{app:gaussproof}.

In order for this to be useful we need to truncate the series Eq.~\eqref{eq:gaussexp} at some finite order $L$. This leads to an approximate kernel function
\begin{equation}
\label{eq:gaussofn}
K_{GL}(\sigma,\omega,\Lambda) = K_G(\sigma,\omega,\Lambda)-R_L(\sigma,\omega,\Lambda)\;,
\end{equation}
where we have defined $R_L$ to be the approximation error. We note in passing that such truncated expansions of the kernel function are routinely used to perform reasonably inversions of the Lorentzian kernel by neglecting the error term $R_L$ as a way of performing a regularization to the ill-posed problem~\cite{Andreasi2005}. The final approximate integral transform $\Phi_{GL}(\omega)$ is then obtained as
\begin{equation}
\label{eq:tgit}
\Phi_{GL}(\omega) = \sum_{k=0}^L c_k\left(\Lambda,\sigma\right) \langle\Psi_G\lvert W_Q^k\rvert\Psi_G\rangle\;.
\end{equation}
As described in Sec.~\ref{sec:intro}, the approximation error contributes to the total variation Eq.~\eqref{eq:totvar} similarly to the approximation error of the time-evolution operator for the simpler Fejer transform. As we did in Sec.~\ref{sec:fejer} we can decompose the total variation as
\begin{equation}
\begin{split}
\delta_V(\Phi_G,\widetilde{\Phi_{GL}})&\leq\delta_V(\Phi_G,\Phi_{GL})+\delta_V(\Phi_{GL},\widetilde{\Phi_{GL}})\\
&\leq R_L(\sigma,\omega,\Lambda)+\delta_V(\Phi_{GL},\widetilde{\Phi_{GL}})\;,
\end{split}
\end{equation}
where $\Phi_G$ is the exact GIT, $\Phi_{GL}$ the approximate integral transform obtained by truncating the series in Eq.~\eqref{eq:gaussexp} at order $L$ and $\widetilde{\Phi_{GL}}$ its finite population estimator.
As we did for the Fejer kernel above, we will now require that both error terms to be less than $\beta/2$ with confidence $\eta_\beta$. In order to bound the total statistical error of the finite population estimator $\widetilde{\Phi_{GL}}$ of the GIT in Eq.~\eqref{eq:tgit} , and assuming for simplicity the same number of measurements for each one of the $L$ expectation values in the expansion, we can take a number of samples given by
\begin{equation}
\begin{split}
N_S&=2L\log\left(\frac{2}{\eta_\beta}\right)\max_{k=\{0,\dots,L\}}\left(L\frac{|c_k|}{\beta}\right)^2\\
&\lesssim2L^3\left(1+\frac{2.2}{\beta}\right)^2\log\left(\frac{2}{\eta_\beta}\right)\;,
\end{split}
\end{equation}
where we used the upperbond on Eq.~\eqref{eq:coeffbound} on $c_k$ obtained in Appendix~\ref{app:gaussproof}. Note that, for technical reasons explained in Appendix~\ref{app:gaussproof}, this is valid only after rescaling the operator $\hat{O}$ by a factor of 2. Since it is possible to find an appropriate bound also in the general case, we do not correspondingly rescale the resolution $\Delta$ here.

The rest of this section will be dedicated to determine an appropriate value for $L$ to ensure $R_L\leq\beta/2$.

In order to find optimal truncation schemes it is now convenient to distinguish between two different situation depending on the desired value $\beta$ as a function of $\Sigma$ and the resolution $\Delta$. More precisely, if we define two critical values $\beta_U$ and $\beta_L$ as follows
\begin{equation}
\beta_L = \frac{1}{\Sigma}\exp\left(-\frac{1}{\Delta^2}\right)\quad\beta_U=\frac{1}{\Delta}\sqrt{\frac{\log(1/\Sigma)}{2}}
\end{equation}
we will try to optimally truncate the polynomial expansion in Eq.~\eqref{eq:gaussexp} in two regimes: the asymptotic regime $\beta\leq\beta_L$ and and intermediate regime where the target accuracy satisfies $\beta_L\leq\beta\leq\beta_U$. Note that the convention we chose in Sec.~\ref{sec:cmp} is compatible with the latter.

As we show in detail in Appendix~\ref{app:gaussproof} (see Eq.~\eqref{eq:asybound} and Eq.~\eqref{eq:intbound}), we can ensure a truncation error $R_L\leq\beta/2$ by choosing the maximum order $L$ according to
\begin{itemize}
    \item in the asymptotic regime $\beta\leq\beta_L$ we need
\begin{equation}
\;\;\;\;\;\;L_{\text{asy}}\!=\!\left\lceil\frac{2e}{\Delta^2}\!\log\left(\frac{1}{\Sigma}\right)\!+\!g_a\!\left(\frac{\Delta^2}{e}\frac{\log\left(6.8/\beta\right)}{\log\left(1/\Sigma\right)}\right)\right\rceil\!-\!2\;,
\end{equation}
where for convenience we have introduced the function $g_a(x) = x/W\left(x\right)$ with $W$ is the Lambert W-function~\cite{Corless1996} (see also Appendix~\ref{app:gaussproof} for details).
    \item in the intermediate regime $\beta_L\leq\beta\leq\beta_U$ we need
\begin{equation}
\label{eq:lintval}
\;\;\;\;\;\;L_{\text{int}}\!=\!\left\lceil\frac{\alpha_1}{\Delta}\sqrt{\log\left(\frac{1}{\Sigma}\right)g_i\left(\frac{\alpha_2}{\Delta\beta}\log\left(\frac{1}{\Sigma}\right)\right)}\;\right\rceil\!-\!1\;,
\end{equation}
with $\alpha_1\lesssim2.93$, $\alpha_2\lesssim4.14$ while the function $g_i$ is
\begin{equation}
g_i(x)=\log\left(x\right)-\frac{1}{4}\log\left(\log\left(x^2\right)\right)\;.
\end{equation}
\end{itemize}
As apparent from the definition of the truncated GIT $\Phi_{GL}$ in Eq.~\eqref{eq:tgit}, this is the maximum required number of invocations to the qubiterate $W_Q$ in a single run since the $L$ expectation values can be computed in parallel. In order to have a better understanding of these results, and connect to the discussion in Sec.~\ref{sec:cmp}, we can write these estimates in terms of asymptotic scaling as
\begin{equation}
L_{\text{asy}} = \mathcal{O}\left(\frac{1}{\Delta^2}\log\left(\frac{1}{\Sigma}\right)+\frac{\log\left(1/\beta\right)}{\log\left(\log\left(1/\beta\right)\right)}\right)\;,
\end{equation}
for the regime with $\beta\leq\beta_L$, in the second regime case with $\beta_L\leq\beta\leq\beta_U$ we find instead
\begin{equation}
\label{eq:lintbound}
L_{\text{int}}=\mathcal{O}\left(\frac{1}{\Delta}\sqrt{\log\left(\frac{1}{\Sigma}\right)\log\left(\frac{1}{\Delta\beta}\log\left(\frac{1}{\Sigma}\right)\right)}\right)\;.
\end{equation}

Note that in applications of the GIT scheme, the concrete values for the truncation order $L$ provided above can be much more useful than the looser bounds Eq.~\eqref{eq:lintbound}. Finally note that, as mentioned in Sec.~\ref{sec:cmp}, the asymptotic regime $\beta\leq\beta_L$ is possibly not directly relevant for the approximate estimation of observables of the form Eq.~\eqref{eq:obsofs}, but could still be helpful in different scenarios. The same argument holds for the ability of the TSA method of Ref.~\cite{Somma2019} to achieve $\Sigma=0$ directly.

\section{Summary and Conclusions}
\label{sec:conclusions}

In this work we have studied a family of quantum algorithms for the approximate estimation of the spectral density operator using ideas from integral transform methods and applied it to the problem of estimating, with bounded errors, the dynamical response function $S(\omega)$ from linear response theory. In particular, we find it useful to consider an integral transform defined by a Gaussian kernel, the Gaussian Integral Transform (GIT). This is in line with the success enjoyed by another integral transform whose kernel is a representation of the delta function: the Lorentz Integral Transform (LIT)~\cite{Efros1994}.

Recently, Somma introduced an algorithm to evaluate multiple eigenvalues based on a time series analysis~\cite{Somma2019}. We show that this technique can be understood in the general framework of integral transform methods introduced here. By comparing it with our GIT we found a quadratic improvement in the regime of interest for the response function approximation problem we are interested in. Notably, our scheme also uses potentially much simpler unitary operations as it completely avoids the need to simulate time evolution under an hamiltonian. This will be important in applications of the GIT based algorithm on NISQ devices. To help implementations of the method, together with a pseudocode implementation of these algorithms in Appendix~\ref{sec:pcode}, we also provide concrete values for the constant factors of all the quantities needed in the practical design of the algorithm.

A possible extension of our algorithm for applications in future fault-tolerant devices is reducing the sample complexity by employing techniques like the method of Ref.~\cite{Knill07} (which uses QPE and Amplitude Amplification) to estimate the expectation values in Eq.~\eqref{eq:exptofk} at the expense of longer circuit depths. Another interesting possibility is to use either Quantum Signal Processing~\cite{Low2019} or the LCU method~\cite{Childs2017} to implement directly the approximate spectral density
\begin{equation}
\hat{\rho}_K(\omega)=K(\omega,\hat{O}) = \sum_{k}^\Gamma K(\omega,O_k)\rvert k\rangle\langle k\lvert\;.
\end{equation}
This would allow, together with Amplitude Amplification, to selectively prepare final states of scattering processes within a pre-determined energy window allowing the application of the algorithm proposed in Ref.~\cite{RoggeroLR} to study rare processes. In such applications the algorithm ceases to be deterministic and a detailed analysis of the failure probability would be needed.

The same strategy can of course be used as a near optimal state preparation scheme similar in many ways to the one recently proposed in Ref.~\cite{Lin20}. Finally, the general framework introduced in this work, and the accuracy metrics defined in Sec.~\ref{sec:intro}, could prove useful to devise alternative approximation schemes based on integral transforms. The interesting question of whether the Gaussian provides the optimal integral kernel for these approximation is left for future work.

After the completion of this manuscript we became aware of a recent similar work by Rall~\cite{rall2020quantum} where an interesting construction for a polynomial representation of a window function was proposed. As we show in Appendix~\ref{sec:patrick} one can use this result to obtain an algorithm for approximating the spectral density with a query complexity $\mathcal{O}(\frac{1}{\Delta}\log\left(\frac{1}{\Sigma\Delta}\right)$. This is an improvement over the scaling of the TSA method by Somma~\cite{Somma2019} but not quite as efficient as the GIT-based method proposed here.

\begin{acknowledgments}

I want to thank M. Savage for his continued support during the preparation of this manuscript, and N. Wiebe for useful comments. This work was supported by the U.S. Department of Energy, Office of Science, Office of Advanced Scientific Computing Research (ASCR) quantum algorithm teams program, under field work proposal number ERKJ333 and by the Institute for Nuclear Theory under U.S. Department of Energy grant No. DE-FG02-00ER41132.
\end{acknowledgments}

%

\appendix
\section{Error bound for frequency observables}
\label{app:proof_obs_bound}

As in the main text we start with a response function
\begin{equation}
\label{eq:sexp}
S(\omega) = \sum_k^{\Gamma}\alpha_k\delta(\omega-\omega_k)\quad\sum_k^{\Gamma}\alpha_k=1\;,
\end{equation}
where we also have $\alpha_k>0$. This decomposition fallows directly from the spectral representation reported in Eq.~\eqref{eq:resp} of the main text. We also define an observable $Q$ which generalizes sum-ruls as the integral
\begin{equation}
Q(S,f) = \int_{-1}^1d \omega S(\omega)f(\omega)\;.
\end{equation}
If we use a $(\Sigma,\Delta,\beta,\eta_\beta)$-approximation $\widetilde{\Phi}(\omega)$ to the response $S(\omega)$ obtained using a $\beta$-approximate estimator for the integral transform $\Phi(\omega)$, we want to find an upperbound for the total error
\begin{equation}
\delta_Q(\widetilde{\Phi}) = \left|Q(S,f)-Q(\widetilde{\Phi},f)\right|\;,
\end{equation}
where the approximate observables is expressed as
\begin{equation}
\label{eq:sampq}
Q(\widetilde{\Phi},f) = \sumint d\nu \widetilde{\Phi}(\nu) \;.
\end{equation}

Using the triangle inequality we find
\begin{equation}
\label{eq:dqapp}
\begin{split}
\delta_Q(\widetilde{\Phi}) &=\left|\int_{-1}^1d\omega S(\omega)f(\omega)-\sumint d\nu\widetilde{\Phi}(\nu)f(\nu)\right|\\
&\leq\left|\int_{-1}^1d\omega S(\omega)f(\omega)-\sumint d\nu\Phi(\nu)f(\nu)\right|\\
&+\left|\sumint d\nu\left(\Phi(\nu)-\widetilde{\Phi}(\nu)\right)f(\nu)\right|\\
&=\delta_Q(\Phi)+\left|\sumint d\nu\left(\Phi(\nu)-\widetilde{\Phi}(\nu)\right)f(\nu)\right|\;.
\end{split}
\end{equation}
In order to find a bound for the first term note that, thanks to the spectral representation Eq.~\eqref{eq:sexp}, we can decompose the total integral transform in the sequence
\begin{equation}
\Phi(\omega)=\sum_k^\Gamma\alpha_k\Phi_k(\omega)\;,
\end{equation}
with $\Phi_K$ the transform of a single peaked response $S_k(\omega)=\delta(\omega-\omega_k)$. We can therefore write
\begin{equation}
\begin{split}
Q(S,f) &= \sum_k^\Gamma\alpha_k\int_{-1}^1d\omega S_k(\omega) f(\omega)=\sum_k^\Gamma\alpha_kf(\omega_k)\\
\end{split}
\end{equation}
while for the integral transform approximator
\begin{equation}
\begin{split}
Q(\Phi,f) &= \sum_k^\Gamma\alpha_k\sumint d\nu\Phi_k(\nu)f(\nu)\\
&= \sum_k^\Gamma\alpha_k\sumint d\nu\!\int\!\! d\omega K(\nu,\omega)S(\omega)f(\nu)\\
&= \sum_k^\Gamma\alpha_k\sumint d\nu K(\nu,\omega_k)f(\nu)\;,
\end{split}
\end{equation}
where in the last line we performed the frequency integral using the decomposition Eq.~\eqref{eq:sexp}.
Using the definition Eq.~\eqref{eq:sigma_approx} of a {\it $\Sigma$-accurate} kernel with {\it resolution $\Delta$} we can find a bound for the first term in Eq.~\eqref{eq:dqapp} as follows
\begin{equation*}
\begin{split}
&\delta_Q(\Phi)=\left|\sum_k^\Gamma\alpha_k\left(f(\omega_k)-\sumint d\nu K(\nu,\omega_k)f(\nu) \right)\right|\\
&\leq\left|\sum_k^\Gamma\alpha_k\left( f(\omega_k)-\sumint_{\omega_k-\Delta}^{\omega_k+\Delta} d\nu K(\nu,\omega_k)f(\nu)\right)\right|\\
&+\left|\sum_k^\Gamma\alpha_k\left(\sumint_{\omega_k+\Delta} \!\!\!d\nu K(\nu,\omega_k)f(\nu)-\sumint^{\omega_k-\Delta}\!\!\! d\nu K(\nu,\omega_k)f(\nu)\right)\right|\\
&\leq \left|\sum_k^\Gamma\alpha_k\left(f(\omega_k)-\sumint_{\omega_k-\Delta}^{\omega_k+\Delta} \!\!\!d\nu K(\nu,\omega_k)f(\nu)\right)\right|+f_{max}\Sigma\\
&\leq \left|\sum_k^\Gamma\alpha_kf(\omega_k)\left(1-\sumint_{\omega_k-\Delta}^{\omega_k+\Delta} \!\!\!d\nu K(\nu,\omega_k)\right)\right|\\
&+\left|\sum_k^\Gamma\alpha_k\sumint_{\omega_k-\Delta}^{\omega_k+\Delta} \!\!\!d\nu K(\nu,\omega_k)\left(f(\nu)-f(\omega_k)\right)\right|+f_{max}\Sigma\\
&\leq \left|\sum_k^\Gamma\alpha_k\sumint_{\omega_k-\Delta}^{\omega_k+\Delta} \!\!\!d\nu K(\nu,\omega_k)\left(f(\nu)-f(\omega_k)\right)\right|+2f_{max}\Sigma\\
&\leq f^\Delta_{max}\left( \sup_{\omega\in[-1,1]}\left|\sumint_{\omega-\Delta}^{\omega+\Delta} \!\!\!d\nu K(\nu,\omega)\right|\right) + 2f_{max}\Sigma
\end{split}
\end{equation*}
with $f_{max}\geq|f(\omega)|$ for all $\omega\in[-1,1]$ and
\begin{equation}
f^{\Delta}_{max} = \sup_{\omega\in[-1,1]} \sup_{x\in[-\Delta,\Delta]} \left|f(\omega+x)-f(\omega)\right|\;.
\end{equation}

Finally, the second term in Eq.~\eqref{eq:dqapp} is bounded as
\begin{equation}
\left|\sumint d\nu\left(\Phi(\nu)-\widetilde{\Phi}(\nu)\right)f(\nu)\right|\leq\beta\int_{-1}^1d\omega\left| f(\omega)\right|
\end{equation}
Bringing all together, we can finally prove the upperbound
\begin{equation}
\delta_Q\leq f^\Delta_{max} + 2f_{max}\Sigma+\beta \int_{-1}^1d\omega \left| f(\omega)\right| 
\end{equation}

\section{Pseudocode implementation}
\label{sec:pcode}

We present here a pseudocode implementation for the spectral density estimation algorithms we discuss in the main text. The goal of our algorithm is to return a $(\Sigma,\Delta,\beta,\eta_\beta)$-approximation $\widetilde{\Phi_k}(\nu)$ to the response function $S(\omega)=\langle\Psi\lvert\hat{\rho}(\omega)\rvert\Psi\rangle$ at a single frequency point $\nu$. This is reasonable since we might not want to estimate $\widetilde{\Phi_k}(\nu)$ on a whole grid composed by the maximal number $\mathcal{O}(1/\Delta)$ of frequency points (as done instead in~\cite{Somma2019}). 

This is however not possible with the Fejer-based method of Section.~\ref{sec:fejer} since the transformed frequencies $\eta$ are {\it sampled} from the distribution $\widetilde{\Phi_k}(\nu)$ instead. For this reason we provide two independent implementations. 

For the Fejer-based strategies we use Algorithm~\ref{alg:1} with
\begin{equation}
M=\mathcal{O}\left(\frac{1}{\Delta\Sigma}\right)\quad N_S=\mathcal{O}\left(\frac{1}{\beta^2}\log\left(\frac{1}{\eta_\beta}\right)\right)\;,
\end{equation}
while $V\equiv\exp(-i2\pi\hat{O})$ for the time dependent method and $V=W_Q$ for the qubitization based method.

\begin{algorithm}[H]
   \caption{Fejer-based approximator\label{alg:1}}
    \begin{algorithmic}[1]
    \State given integers $M=2^m$ and $N_S$
    \For{$i = 1$ to $N_S$}
        \State prepare target state $\ket{\Psi}$
        \State apply QPE with unitary $V$ and maximum order $V^{M/2}$
        \State measure $m$ qubits in ancilla register in frequency $\nu_i$
        \State add result to frequency histogram
    \EndFor
    \State return frequency histogram
\end{algorithmic}
\end{algorithm}

In the case of either the TSA algorithm or the GIT-based method we can use Algorithm~\ref{alg:2} instead with a maximum order $M$ given by
\begin{equation}
M_{TSA}=\mathcal{O}\left(\frac{1}{\Delta}\log\left(\frac{1}{\beta}\right)^2\right)
\end{equation}
for the TSA algorithm of~\cite{Somma2019}, while for the GIT
\begin{equation}
M_{GIT}=\mathcal{O}\left(\frac{1}{\Delta}\sqrt{\log\left(\frac{1}{\Sigma}\right)\log\left(\frac{1}{\Delta\beta}\log\left(\frac{1}{\Sigma}\right)\right)}\right)\;,
\end{equation}
together with a number of samples per order scaling as
\begin{equation}
N=\mathcal{O}\left(\frac{M^2}{\beta^2}\log\left(\frac{1}{\eta_\beta}\right)\right)\;.
\end{equation}

\begin{algorithm}[H]
   \caption{Orthogonal polynomial-based approximator\label{alg:2}}
    \begin{algorithmic}[1]
    \State given integers $M$ and $N_S=M\times N$
    \For{$k = 1$ to $M$}
        \For{$i = 1$ to $N$}
            \State prepare target state $\ket{\Psi}$
            \State measure expectation value $v_k=\langle\Psi\lvert V^k\rvert\Psi\rangle$
        \EndFor
        \State store estimator of $v_k$ with error $\mathcal{O}(1/\sqrt{N})$
    \EndFor
    \State compute expansion coefficients $\vec{c}(\nu)$ corresponding to the integral transform being evaluated at target frequency $\nu$
    \State return $\widetilde{\Phi_k}(\nu)=\vec{c}(\nu)\cdot\vec{v}$ 
\end{algorithmic}
\end{algorithm}

Finally, if we want the transform $\widetilde{\Phi_k}(\nu)$ at all the $\mathcal{O}(1/\Delta)$ frequency points while keeping the {\it total} error $\beta$ (as considered in~\cite{Somma2019}) we will need instead
\begin{equation}
M_{TSA}=\mathcal{O}\left(\frac{1}{\Delta}\log\left(\frac{1}{\Delta\beta}\right)^2\right)
\end{equation}
and
\begin{equation}
M_{GIT}=\mathcal{O}\left(\frac{1}{\Delta}\sqrt{\log\left(\frac{1}{\Sigma}\right)\log\left(\frac{1}{\Delta^2\beta}\log\left(\frac{1}{\Sigma}\right)\right)}\right)\;,
\end{equation}
respectively. In this case we see that the logarithmic term for TSA also contain the resolution scale $\Delta$.

\section{Error analysis for faulty implementation of Fejer}
\label{app:proof_faulty_fejer}
Assume we have an approximation to the phase kick-back part of the QPE algorithm (the application of the controlled-$U$ operations) which satisfies
\begin{equation}
\|\widetilde{V}_{\text{PKB}}-V_{\text{PKB}}\|\leq\delta\;.
\end{equation}
Define $\ket{\Phi_A}\!=\!V_{\text{PKB}}\ket{\Psi}$, without loss of generality we have
\begin{equation}
\ket{\Phi_B}=\widetilde{V}_{\text{PKB}}\ket{\Psi}=\cos(\alpha)\ket{\Phi_A}+\sin(\alpha)\ket{\xi}
\end{equation}
with $\langle\Phi_A\vert\xi\rangle=0$. If we introduce the density matrices $\rho=\rvert\Phi_A\rangle\langle\Phi_A\lvert$ and $\sigma=\rvert\Phi_B\rangle\langle\Phi_B\lvert$, we can now write
\begin{equation*}
\sqrt{2(1-\cos(\alpha))}\|\ket{\Phi_A}-\ket{\Phi_B}\|_2\leq\|\widetilde{V}_{\text{PKB}}-V_{\text{PKB}}\|\leq\delta
\end{equation*}
then $\cos(\alpha)\geq1-\delta^2/2$, but also
\begin{equation}
D(\rho,\sigma)=\frac{1}{2}Tr[|\rho-\sigma|]=|\sin(\alpha)|\leq\delta\sqrt{1-\frac{\delta^2}{4}}\;.
\end{equation}
We can write the transform at location $\sigma_q=2q/N-1$ as
\begin{equation}
\begin{split}
\Phi_F(\sigma_q,\Delta,N)&= \Phi_F(\sigma_q)=Tr\left[\Pi_q U_{QFT}^\dagger \rho U_{QFT}\right]\;,
\end{split}
\end{equation}
where $\Pi_q=\rvert q\rangle\langle q\lvert$ and $U_{QFT}$ the unitary implementing the Quantum Fourier Transform on the ancilla register. A similar expression holds for the faulty density matrix $\sigma$. We now have for any $0\leq q<N$ that
\begin{equation}
\begin{split}
\delta_V(\Phi_F,\widetilde{\Phi_F})&\leq\sup_{\|\widetilde{V}_{\text{PKB}}-V_{\text{PKB}}\|\leq\delta} \left|\Phi_F(\sigma_q)-\widetilde{\Phi_F}(\sigma_q)\right|\\
&=\left|Tr\left[\Pi_q U_{QFT}^\dagger (\rho-\sigma)U_{QFT}\right]\right|\\
&\leq\frac{1}{2}Tr\left[|\rho-\sigma|\right]\leq\delta\sqrt{1-\frac{\delta^2}{4}}\leq\delta\;.
\end{split}
\end{equation}
Furthermore, since $V_{\text{PKB}}$ is a product of $n=log_2(N)$ controlled time evolution unitaries
\begin{equation}
V_{\text{PKB}} = \prod_{k=0}^{n-1} U(t=2\pi2^k)\;,
\end{equation}
we have by the union bound that
\begin{equation}
\begin{split}
\|\widetilde{V}_{\text{PKB}}&-V_{\text{PKB}}\|\leq \sum_{k=0}^{n-1}\|U(2\pi2^k)-\widetilde{U}(2\pi2^k)\|\\
&\leq n \max_{0\leq<k<n}\delta_t(2\pi2^k)
\end{split}
\end{equation}
In the last equation $\delta_t(\tau)$ is the approximation error of the time evolution unitary for total time $t=\tau$. If we choose all approximation errors to be the same $\delta_t(\tau)=\delta_t$ then we find 
\begin{equation}
\delta_V(\Phi_F,\widetilde{\Phi_F}) \leq\log_2(N)\delta_t\;.
\end{equation}

\section{Chebychev expansion of the gaussian kernel}
\label{app:gaussproof}
Using an expansion in Chebyschev polynomials, we can express the Gaussian function as
\begin{equation}
\exp\left(-\frac{x^2}{2\Lambda^2}\right)=\sum_{n=0}^L a_n(\Lambda)T_n(x)+r_L(x,\Lambda)\;,
\end{equation}
where $r_L(\Lambda)$ indicates the truncation error and the coefficients are given by (see Eq.(4) of Ref.~\cite{Tausch09a})
\begin{equation}
a_n = \bigg\{\begin{matrix}
\frac{\gamma_n}{\sqrt{2\pi}\Lambda}i^{\frac{n}{2}}\exp\left(-\frac{1}{4\Delta^2}\right)J_{n/2}\left(\frac{i}{4\Lambda^2}\right)&\text{for even {\it n}} \\
0&\text{for odd {\it n}}
\end{matrix}\;,
\end{equation}
with $\gamma_0=1$ and $\gamma_{n>0}=2$ and $J_n$ the Bessel function of order $n$. Before discussing bounds on the magnitude of the truncation error, we want to first discuss how the kernel function $K_G(\sigma,\omega,\Lambda)$ can be generated using the expansion above. First note that we can write the truncated kernel function as in Eq.~\eqref{eq:gaussexp} of the main text
\begin{equation}
\label{eq:gexpapp}
K_{GL}(\sigma,\omega,\Lambda) =\frac{1}{\sqrt{2\pi}\Lambda} \sum_{k=0}^L a_k\left(\frac{\Lambda}{2}\right)T_k\left(\frac{\omega-\sigma}{2}\right)\;.
\end{equation}
Since $T_k\left(\frac{\omega-\sigma}{2}\right)$ is a polynomial of degree $k\leq L$ we have
\begin{equation}
T_k\left(\frac{\omega-\sigma}{2}\right) = \sum_{j=0}^L b_{jk}(\sigma) T_j(\omega)\;,
\end{equation}
where the expansion coefficients are given by
\begin{equation}
\begin{split}
b_{jk}(\sigma) &= \frac{\gamma_j}{\pi} \int_{-1}^1 \frac{dx}{\sqrt{1-x^2}} T_k\left(\frac{x-\sigma}{2}\right)T_j(x)\\
&= \frac{\gamma_j}{L}\sum_{m=0}^{L-1}T_k\left(\frac{x_m-\sigma}{2}\right)T_j(x_m)\;.
\end{split}
\end{equation}
In the second line we used Gauss-Chebyschev quadrature and $x_m=\cos\left(\pi\frac{2m-1}{2L}\right)$ the Chebyshev nodes (this is similar to the strategy used in Ref.~\cite{Tausch09a}). Using this representation we can rewrite the kernel function as
\begin{equation}
K_{GL}(\sigma,\omega,\Lambda) = \sum_{j=0}^L c_j\left(\Lambda,\sigma\right)T_j\left(\omega\right)
\end{equation}
where the new expansion coefficients are given by 
\begin{equation}
\label{eq:ccoeff}
c_j = \frac{\gamma_j}{\sqrt{2\pi}\Lambda L}\sum_{m=0}^{L-1}\sum_{k=0}^La_k\left(\frac{\Lambda}{2}\right)T_k\left(\frac{x_m-\sigma}{2}\right)T_j(x_m).
\end{equation}

\subsection{Bound of expansion coefficients}

We can bound the magnitude of $c_j$ as follows
\begin{equation}
\begin{split}
\left|c_j\right|\! &= \!\left|\frac{\gamma_j}{\sqrt{2\pi}\Lambda L}\!\!\sum_{m=0}^{L-1}\!\left(e^{-\frac{(x_m-\sigma)^2}{2\Lambda^2}}\!\!-r_L\left(x_m,\frac{\Lambda}{2}\right)\right)T_j(x_m)\right|\\
&\leq\frac{\gamma_j}{\sqrt{2\pi}\Lambda }r_L\left(\frac{\Lambda}{2}\right)+\frac{\gamma_j}{\sqrt{2\pi}\Lambda L}\sum_{m=0}^{L-1}e^{-\frac{(x_m-\sigma)^2}{2\Lambda^2}}\\
&\coloneqq\gamma_jR_L\left(\Lambda\right)+\Omega_j\;,
\end{split}
\end{equation}
with $R_L(\Lambda)$ the truncation error of the kernel function (cf. Eq.~\eqref{eq:gaussofn}). For the second term we can use
\begin{equation}
\begin{split}
\Omega_j&\leq\frac{\gamma_j}{\sqrt{2\pi}\Lambda L}\int_{0}^Ldx\exp\left(-\frac{(\cos\left(\pi\frac{2x-1}{2L}\right)-\sigma)^2}{2\Lambda^2}\right)\\
&=\frac{\gamma_j}{\sqrt{2\pi^3}\Lambda}\int_{0}^{\pi}dy\exp\left(-\frac{(\cos\left(y-\frac{\pi}{2L}\right)-\sigma)^2}{2\Lambda^2}\right)\;.
\end{split}
\end{equation}
The integral approximately measures the number of of Chebyshev nodes within the envelope of the gaussian kernel centered at $\sigma$. Since these nodes cluster near the edges of the interval $[-1,1]$, we can obtain coefficients with a smaller maximum magnitude by rescaling the energy spectrum into a smaller interval and considering transformed variables $\sigma$ in the same restricted interval. As we mention in the main text we work here with the assumption that $\omega\in[-1/2,1/2]$ and the same for $\sigma$.

Now we use the following bound for the cosine term
\begin{equation}
\left(\cos\left(y-\frac{\pi}{2L}\right)-\sigma\right)^2 \geq \left(\cos\left(y\right)-\sigma\right)^2 -\left(\frac{\pi}{L}\right)^2\;,
\end{equation}
to simplify the integrand above and obtain then
\begin{equation}
\begin{split}
\Omega_j&\leq\frac{\gamma_j}{\sqrt{2\pi^3}\Lambda}e^{\frac{\pi^2}{2L^2\Lambda^2}}\int_{0}^{\pi}\!\!dy\exp\left(-\frac{(\cos\left(y\right)-\sigma)^2}{2\Lambda^2}\right)\\
=&\frac{\gamma_j}{\sqrt{2\pi^3}\Lambda}e^{\frac{\pi^2}{2L^2\Lambda^2}}\int_{-1}^1\frac{dx}{\sqrt{1-x^2}}\exp\left(-\frac{(x-\sigma)^2}{2\Lambda^2}\right)\;.
\end{split}
\end{equation}
Finally using the fact that we rescaled the energies so that $\sigma\in[-1/2,1/2]$, we can bound the integral by
\begin{equation}
\begin{split}
\Omega_j&\!\leq\!\frac{\gamma_j}{\sqrt{2\pi^3}\Lambda}e^{\frac{\pi^2}{2L^2\Lambda^2}}\!\!\int_{-1}^1\!\!\frac{dx}{\sqrt{1-x^2}}\exp\left(-\frac{(x-1/2)^2}{2\Lambda^2}\right)\\
&\leq \frac{\gamma_j}{\sqrt{2\pi^3}\Lambda}\exp\left(\frac{\pi^2}{2L^2\Lambda^2}\right) \left(2.5\Lambda\right)\;,
\end{split}
\end{equation}
where the constant factor in the second line was obtained numerically. In summary, we found the following bound
\begin{equation}
\label{eq:coeffbound}
\begin{split}
\left|c_j(\Lambda,\sigma)\right|&\leq \gamma_j \left(R_L(\Lambda) + 0.32\exp\left(\frac{\pi^2}{2L^2\Lambda^2}\right)\right)\\
&\leq 2\left(R_L(\Lambda) + 1.1\right)
\end{split}
\end{equation}
where we anticipated the result $L\Lambda>2$ that will be proved in the next two section. 

\subsection{Bound on truncation error}

We turn now to providing upperbounds for the error terms $r_L(\Lambda)$ and $R_L(\Lambda)$. Using the result from Tausch and Weckiewicz~\cite{Tausch09a} we can bound the magnitude of the expansion coefficients as
\begin{equation}
\left|a_n(\Lambda)\right|\leq 2\Lambda\sqrt{\pi}\exp\left(-(n+1)\kappa\left((n+1)2\Lambda^2\right)\right)\;,
\end{equation}
where the auxiliary function $\kappa(x)$ is given by
\begin{equation}
\label{eq:kappa}
\begin{split}
\kappa(x)&=\frac{\log(x+\sqrt{1+x^2})}{2}-\frac{1}{4x}\frac{\left(x-1+\sqrt{1+x^2}\right)^2}{x+\sqrt{1+x^2}}\;.
\end{split}
\end{equation}

The total error $r_L(\Lambda)$ can then be bounded summing a geometric series\footnote{Note the missing factor of 2 from Eq.~(11) of~\cite{Tausch09a}}, the result is
\begin{equation}
\label{eq:rNbound0}
\begin{split}
\left|r_L(\Lambda)\right|&=\left|\sum_{n=L+1}^\infty a_n(\Lambda)T_n(x)\right|\leq\sum_{n=L+1}^\infty\left| a_n(\Lambda)\right| \\
&\leq 2\Lambda \sqrt{\pi}\frac{\exp\left(-L'\kappa(2L'\Lambda^2)\right)}{1-\exp\left(-\kappa(2L'\Lambda^2)\right)}
\end{split}
\end{equation}
with $L'=L+2$ for $L$ even and $L'=L+3$ for $L$ odd. We can obtain a simpler upper-bound by first using the fact that for $x>1$ we can bound $\kappa(x)$ with
\begin{equation}
\kappa(x)\geq \frac{1}{2}\left(\log(2x)-1\right)\;,
\end{equation}
and then using the monotonicity of the denominator in Eq.~\eqref{eq:rNbound0} to find, for $2L'\Lambda^2\geq1$, the bound
\begin{equation}
\left|r_L(\Lambda)\right|\leq \frac{2\Lambda \sqrt{\pi}}{1-\exp\left(-\kappa(1)\right)}\left(\frac{e}{4L'\Lambda^2}\right)^{\frac{L'}{2}}\;.
\end{equation}
Using this result we find the total error $R_N(\sigma,\Lambda)$ in the gaussian transform Eq.~\eqref{eq:gaussofn} to be bounded as
\begin{equation}
\label{eq:fullboundasy}
\begin{split}
\left|R_L(\sigma,\Lambda)\right| &= \frac{1}{\sqrt{2\pi}\Lambda}\left|r_L\left(\frac{\Lambda}{2}\right)\right|\\
&\leq\frac{1}{\sqrt{2}}\frac{1}{1-\exp\left(-\kappa(1)\right)}\left(\frac{e}{L'\Lambda^2}\right)^{\frac{L'}{2}}\\
&\lesssim3.4\left(\frac{e}{L'\Lambda^2}\right)^{\frac{L'}{2}}\;,
\end{split}
\end{equation}
valid in the asymptotic regime $L'\geq2/\Lambda^2$. In order to guarantee a truncation error of at most $\epsilon_R$ we now need
\begin{equation}
\label{eq:ineqA}
\epsilon_R\geq3.4\left(\frac{e}{(L+2)\Lambda^2}\right)^{\frac{(L+2)}{2}}
\end{equation}
for the number of repetitions $L$. Note that in this last expression we used the conservative value $L'=L+2$. The inequality in Eq.~\eqref{eq:ineqA} can be solved as
\begin{equation}
\label{eq:asybound}
L\geq\frac{e}{\Lambda^2}-2+\frac{2\Lambda^2}{e}\frac{\log\left(3.4/\epsilon_R\right)}{W\left(\frac{2\Lambda^2}{e}\log\left(3.4/\epsilon_R\right)\right)}
\end{equation}
where $W$ is the Lambert W-function~\cite{Corless1996}. In order to understand the scaling of this expression we can use the less tight bound
\begin{equation}
L\geq\frac{e}{\Lambda^2}+\frac{\log\left(3.4/\epsilon_R\right)}{\log\left(\log\left(3.4/\epsilon_R\right)\right)}-2\;.
\end{equation}
which is usually employed in the literature (see eg.~\cite{Low2017}).

\subsection{Intermediate regime}
We will now provide bounds in the second regime considered in the main text where the upperbound on the order $L$ is the lower limit of validity for Eq.~\eqref{eq:fullboundasy}, namely $0<L'\leq\frac{2}{\Lambda^2}$. In this case there is a minimum error which we can guarantee, the value of which we will determine in this section (see Eq.~\eqref{eq:erminapp}). 
We can start by first noticing that for $0<x\leq1$ we have
\begin{equation}
x\kappa(1)\leq \kappa(x)\leq\frac{x}{4}
\end{equation}
so that we can bound the total error $r_N$ using
\begin{equation}
\begin{split}
\sum_{n=L+1}^\infty\left| a_n(\Lambda)\right|&\leq2\Lambda\sqrt{\pi}\sum_{n=L+1}^\infty e^{-(n+1)\kappa\left((n+1)2\Lambda^2\right)}\\
&\leq2\Lambda\sqrt{\pi}\sum_{n=L+1}^\infty e^{-(x+1)^22\kappa(1)\Lambda^2} \\
&\leq2\Lambda\sqrt{\pi}\int_{L}^\infty dx e^{-(x+1)^22\kappa(1)\Lambda^2}\\
&=\frac{\pi}{\sqrt{2\kappa(1)}}\text{erfc}\left((L+1)\Lambda\sqrt{2\kappa(1)}\right)\;.
\end{split}
\end{equation}
This, in turn, implies the following upper-bound for the error in the transform
\begin{equation}
\!\!\left|R_L(\sigma,\Lambda)\right|\leq\frac{1}{2\Lambda}\sqrt{\frac{\pi}{\kappa(1)}}\text{erfc}\left((L+1)\Lambda\sqrt{\frac{\kappa(1)}{2}}\right).
\end{equation}
This error con be bounded from above using
\begin{equation}
\label{eq:trunc_bound_B}
\left|R_L(\sigma,\Lambda)\right|\leq\frac{1}{\sqrt{2}\Lambda_\kappa^2}\frac{1}{L+1}\exp\left(-(L+1)^2\frac{\Lambda_\kappa^2}{2}\right)
\end{equation}
where we defined $\Lambda_\kappa=\Lambda\sqrt{\kappa(1)}$, and is valid for
\begin{equation}
\label{eq:cond_ir_A}
L\geq\sqrt{\frac{2}{\pi}}\frac{1}{\Lambda_k}-1\;.
\end{equation}

As we did in the general case above, if we want a truncation error of at most $\epsilon_R$ we need
\begin{equation}
\epsilon_R\geq\frac{1}{\sqrt{2}\Lambda_\kappa^2}\frac{1}{L+1}\exp\left(-(L+1)^2\frac{\Lambda_\kappa^2}{2}\right)\;.
\end{equation}
The solution can again be conveniently expressed in terms of the Lambert W-function as
\begin{equation}
L+1\geq\frac{1}{\Lambda_k}\sqrt{\frac{1}{2}W\left(\frac{1}{2\Lambda_\kappa^2\epsilon_R^2}\right)}\;.
\end{equation}
We can now use another result from~\cite{Hoorfar2008}, Theorem 2.1, to find the sufficient condition
\begin{equation}
\label{eq:intbound}
L=\left\lceil\frac{1}{\Lambda}\sqrt{\frac{1}{\kappa(1)}g\left(\frac{1}{\sqrt{2\kappa(1)}\Lambda\epsilon_R}\right)}\;\;\right\rceil-1\;,
\end{equation}
where for convenience we defined the auxiliary function
\begin{equation}
g(x)=\log\left(x\right)-\frac{1}{4}\log\left(\log\left(x^2\right)\right)\;.
\end{equation}
These estimates hold for sufficiently small target errors
\begin{equation}
\Lambda\epsilon_R\leq\frac{1}{\sqrt{2\kappa(1)}e}\approx0.54\;,
\end{equation}
a conditions that ensures that also Eq.~\eqref{eq:cond_ir_A} is satisfied.
We finally note that it is also possible to find a bound on $L$ valid for any value of the target error
\begin{equation}
L=\left\lceil\frac{1}{\Lambda}\sqrt{\frac{2}{\kappa(1)}\log\left(\sqrt{\frac{\pi}{\kappa(1)}}\frac{1}{2\Lambda\epsilon_R}\right)}\right\rceil\;.
\end{equation}

We now need to find the minimum error that can be guaranteed in this intermediate regime, using the upper-bound from Eq.~\eqref{eq:trunc_bound_B} we find
\begin{equation}
\label{eq:erminapp}
\begin{split}
\epsilon_R^{\text{min}} &\leq\frac{1}{\sqrt{2}\kappa(1)}\frac{1}{2+\Lambda^2}\exp\left(-\frac{\kappa(1)}{2}\frac{(2+\Lambda)^2}{\Lambda^2}\right)\\
&\leq\frac{e^{-2\kappa(1)}}{\sqrt{8}\kappa(1)}\exp\left(-\frac{2\kappa(1)}{\Lambda^2}\right)\lesssim\exp\left(-\frac{1}{2\Lambda^2}\right)\;.
\end{split}
\end{equation}
and this is valid for any reasonable value $\Lambda$ (to satisfy Eq.~\eqref{eq:cond_ir_A} the condition is $\Lambda\leq{5}$).

\section{Kernel based on Jackson's theorem}
\label{sec:patrick}
The recent work by Rall~\cite{rall2020quantum} introduced an integral transform kernel based on Jackson's theorem from approximation theory. In this appendix we use the relevant results from Ref.~\cite{rall2020quantum} to construct an approximate integral transform and compare it with the GIT and TSA-based methods described in the main text.
The approximate window function introduced in Ref.~\cite{rall2020quantum} can be used to construct a (normalized) integral kernel as
\begin{equation}
K_J(\sigma,\omega,k,N) = \mathcal{N}_{kN}\omega_{kN}\left(\frac{\sigma-\omega}{2}\right)\;,
\end{equation}
with $\mathcal{N}_{kN}$ a normalization factor and 
\begin{equation}
\omega_{kN}(x) = A_k\left(\frac{4}{5}J_N(x)\right)\;.
\end{equation}
In the expression above, $A_k$ is the amplifying polynomial from Eq.(A5) of Ref.~\cite{rall2020quantum}, while $J_N(x)$ is the Jackson's approximation to the function $g(x)$ defined as:
\begin{equation}
\label{eq:app_gfunc}
g(x) = \left\{\begin{matrix}
-1&x<-\delta\\
-1 + \frac{2}{\delta}\left(x+\delta\right)&-\delta<x\leq0\\
1 -\frac{2}{\delta}x&0>x>\delta\\
-1&x>\delta\\
\end{matrix}
\right.\;,
\end{equation}
for some fixed resolution $\delta$. Note that in this construction we let the approximation interval $[\bar{a},\bar{b}]$ defined in Ref.~\cite{rall2020quantum} shrink to zero. As shown in Ref.~\cite{rall2020quantum}, in order to ensure $J_N$ approximates $g(x)$ with error less than $1/4$ one can take $N=24/\delta$. The order $k$ of $A_k$ controls the final approximation error by ensuring that for $x>\delta$ the final function satisfies
\begin{equation}
A_k\left(\frac{4}{5}J_N(x)\right)\leq\tau\equiv\exp(-k/6)\;.
\end{equation}

The condition for the integral transform to be $\Sigma$-accurate with resolution $\Delta$ can be written as
\begin{equation}
\label{eq:app_sig_cond}
\begin{split}
\sup_{\omega_0\in[-1,1]} &\left(\int_{-1}^{\omega-\Delta} d\sigma K_J(\sigma,\omega_0,k,N) \right. \\
&\left.+ \int_{\omega+\Delta}^1 d\sigma K_J(\sigma,\omega_0,k,N)\right)\leq\Sigma\;,
\end{split}
\end{equation}
or in the more convenient form
\begin{equation}
2\mathcal{N}_{kN} \int_{\Delta/2}^1 dx A_k\left(\frac{4}{5}J_N\left(x\right)\right)\leq\Sigma\;.
\end{equation}

By choosing the resolution in the $g$ function in Eq.~\eqref{eq:app_gfunc} to be $\delta=\Delta/2$, we find that Eq.~\eqref{eq:app_sig_cond} is satisfied for
\begin{equation}
\tau \leq \frac{\Sigma}{2-\Delta}\frac{1}{\mathcal{N}_{kN}}\;.
\end{equation}
The normalization constant can be bounded using
\begin{equation}
1=\int_{-1}^1 dx K_J(x,\omega,k,N) \leq \mathcal{N}_{kN}\left(2\delta+\tau(2-2\delta)\right)\;,
\end{equation}
and this gives the following necessary condition on $\tau$
\begin{equation}
\tau \leq \Sigma\frac{\Delta+\tau(2-\Delta))}{2-\Delta}\Rightarrow \tau \leq \frac{\Sigma}{1-\Sigma}\frac{\Delta}{2-\Delta}
\end{equation}
If we require the approximation to be $\Sigma$-accurate with resolution $\Delta$, the order $d$ of the polynomial representation of the kernel $K_J$ needs to be larger than
\begin{equation}
d_{min}= \frac{288}{\Delta}\log\left(\frac{1-\Sigma}{\Sigma}\frac{2-\Delta}{\Delta}\right)\;.
\end{equation}

The asymptotic cost of using the Jackson kernel for the spectral density approximation is therefore worse than then the GIT-based method presented in the main text. Comparing this result with the TSA method of Ref.~\cite{Somma2019} will however require to find an upperbound on the normalization constant first. We can obtain this by noticing that, in the intervals $[0,\delta/2]$ and $[-\delta/2,0]$, the kernel function can be bounded from below using a linear function while outside of this region the lower bound is zero. We can therefore write for $\tau<5/8$ the following
\begin{equation}
\mathcal{N}_{kN} \leq \frac{1}{\delta}\frac{4}{5-8\tau}\;,
\end{equation}
which recovers the intuition that in general $\mathcal{N}_{kN}$ should scale linearly with the resolution. This shows that the method presented in this appendix has also a better complexity than the TSA algorithm.

\end{document}